\documentclass[showpacs,preprintnumbers,amsmath,amssymb,prb]{revtex4}

\usepackage{graphicx}
\usepackage{dcolumn}
\usepackage{bm}

\begin{document}

\title{Collective excitations in two-dimensional antiferromagnet in strong magnetic field}

\author{A. V. Syromyatnikov}
 \email{syromyat@thd.pnpi.spb.ru}
\affiliation{Petersburg Nuclear Physics Institute, Gatchina, St.\ Petersburg 188300, Russia}

\date{\today}

\begin{abstract}

We discuss spin-$\frac12$ two-dimensional (2D) Heisenberg antiferromagnet (AF) on a square lattice at $T=0$ in strong magnetic field $H$ near its saturation value $H_c$. A perturbation approach is proposed to obtain spectrum of magnons with momenta not very close to AF vector in the leading order in small parameter $(H_c-H)/H_c$. We find that magnons are well-defined
quasi-particles at $H>0.9H_c$ although the damping is quite large near the zone boundary. A characteristic rotonlike local minimum in the spectrum is observed at ${\bf k}=(\pi,0)$ accompanied by decrease of the damping near $(\pi,0)$. The suggested approach can be used in discussion of short-wavelength excitations in other 2D Bose gases of particles or quasi-particles.

\end{abstract}

\pacs{75.10.Jm, 75.50.Ee, 75.40.Gb}

\maketitle

\section{Introduction}

Spin-$\frac12$ two-dimensional (2D) Heisenberg antiferromagnet (AF) on a square lattice has been one of the most attractive theoretical objects in the last two decades because this model
describes parent compounds of high-$T_c$ superconducting cuprates \cite{monous}. Whereas the theory of long-wavelength magnons in quantum square 2D AF is well developed and describes well existing experimental data \cite{monous,chak,christ}, there are some surprising recent experimental findings indicating that the standard theoretical approaches do not work for short-wavelength magnons. Thus, a rotonlike local minimum was observed at small $T$ in the spin-wave dispersion at ${\bf k}=(\pi,0)$ in a number of recent experiments on square spin-$\frac12$ 2D AFs \cite{roton1,roton2,christ,lumsden}. This minimum is not reproduced quantitatively within second order of $1/S$ expansion and phase flux RVB techniques, while numerical computations using Quantum Monte-Carlo and series expansion describe it satisfactorily (see Ref.~\cite{christ} and references therein). The origin of the rotonlike minimum has not been clarified yet. It is attributed to the entanglement of spins on neighboring sites \cite{christ}. Rotonlike minimum in the spectrum of short-wavelength magnons (at ${\bf k}=(\pi,\pi/\sqrt3)$) has been reported recently also in spin-$\frac12$ triangular AF at $H=0$ obtained using $1/S$ expansion \cite{starykh,zhitcher06} and series expansion \cite{zheng_prl,zheng_prb}.

Even more peculiar finding concerning short-wavelength magnons in square spin-$\frac12$ 2D AF was obtained in Ref.~\cite{zhitcher99}. The authors investigated renormalization of the spin-wave spectrum within first order in $1/S$ in strong magnetic field $H$ smaller than the saturation field $H_c$. Due to noncollinearity of sublattices and the field, three-magnon terms appear in the Hamiltonian after transformation of spin operators to bosons. Such terms lead to the loop diagram for the self-energy part $\Sigma(k)$ in the first order in $1/S$ corresponding to processes of magnon decay into two magnons. The spin-wave damping obtained within the first order in $1/S$ was not very large. To discuss decaying processes self-consistently authors went beyond the first order in $1/S$ carrying out a self-consistent calculation of the Green's function (GF) $G(\omega,{\bf k})$ using the dressed GF considering the loop diagram. As a result they obtained that ${\rm Im}G(\omega,{\bf k})$ as a function of $\omega$ at fixed $\bf k$ does not resemble a peak at all in most of the Brillouin zone (BZ) when $0.76H_c<H<H_c$ (the worse situation was at $H\approx0.9H_c$). Thus the authors concluded that magnons are unstable according to decay into two magnons at $H\sim H_c$. Magnons reappeared only at $H>0.99H_c$ due to decreasing of the three-magnon vertex. Meantime within the approach suggested in Ref.~\cite{zhitcher99} only some higher-order $1/S$ corrections are taken into account and it is not clear what is the role of the ignored $1/S$ terms. This question looks very serious because there is no small parameter in $1/S$ expansion for $S=1/2$. On the other hand the unsuccessful attempts to describe rotonlike minimum in the spectrum within second order in $1/S$ at $H=0$ and results of Ref.~\cite{zhitcher99} themselves signify that higher order $1/S$ terms can renormalize the spectrum of short-wavelength magnons noticeably if $S\sim1$.

In view of the above mentioned results it is highly desirable to develop an analytical perturbation theory allowing quantitative discussion of the short-wavelength magnons in quantum 2D AF. The aim of the present paper is to develop such theory in the
parameter $(H_c-H)/H_c\ll1$ for magnons with momenta not very close to AF vector ${\bf k}_0=(\pi,\pi)$. The magnon spectrum is calculated below in the leading order in $(H_c-H)/H_c$. Our results can be tested experimentally on the growing family of AF compounds with small or moderate exchange coupling constant $J$ (i.e., with experimentally achievable $H_c\propto J$) which can be modeled by square spin-$\frac12$ Heisenberg AF \cite{wood,lancaster,coomer}. The approach suggested below can be used in discussion of short-wavelength excitations in other 2D Bose gases of particles or quasi-particles.

The rest of the paper is organized as follows. We discuss the approach in Sec.~\ref{method}. Results of the spectrum calculation are presented in Sec.~\ref{res}. Sec.~\ref{con} contains our conclusion. One appendix is added devoted to diagrams estimations.

\section{Model and technique}
\label{method}

We discuss spin-$\frac12$ Heisenberg AF on the square lattice in magnetic field which Hamiltonian has the form
\begin{equation}
\label{ham}
{\cal H} = \frac12 \sum_{i,j} J_{ij} {\bf S}_i{\bf S}_j + H\sum_i S_i^z.
\end{equation}
We express spin projections via Pauli operators $a_i^\dagger$ and $a_i$ as
$S_i^z = -\frac12 + a_i^\dagger a_i$, $\quad S_i^\dagger = a_i^\dagger$, $\quad S_i^- = a_i$. To treat them as Bose operators one should introduce to Hamiltonian (\ref{ham}) the constraint term $U\sum_i a^\dagger_i a^\dagger_i a_i a_i$, where $U\to\infty$, describing infinite repulsion of particles on the same site \cite{bat84}. As a result Hamiltonian (\ref{ham}) reads as
\begin{eqnarray}
\label{hbg}
{\cal H} 
&=& 
\sum_{\bf k} \left(\epsilon_{\bf k}  - \mu\right) a^\dagger_{\bf k} a_{\bf k}
\nonumber\\
&&{}
+
\frac{1}{2N}
\sum_{{\bf k}_1+{\bf k}_2 = {\bf k}_3+{\bf k}_4}
(J_{{\bf k}_1-{\bf k}_3} + U) a^\dagger_{{\bf k}_1} a^\dagger_{{\bf k}_2} a_{{\bf k}_3} a_{{\bf k}_4},
\end{eqnarray}
where $\epsilon_{\bf k} = \frac12(J_{\bf 0} + J_{\bf k})$, $J_{\bf k}=\sum_i J_{ij}\cos({\bf k R}_{ij})$, ${\bf R}_{ij}$ connects sites $i$ and $j$, $\mu = H_c - H$, $H_c=J_{\bf 0}$ is the saturation field, and $N$ is the number of spins in the lattice. The bare spectrum $\epsilon_{\bf k}$ is quadratic near ${\bf k}_0$.

\begin{figure}
\centering
\includegraphics[scale=0.63]{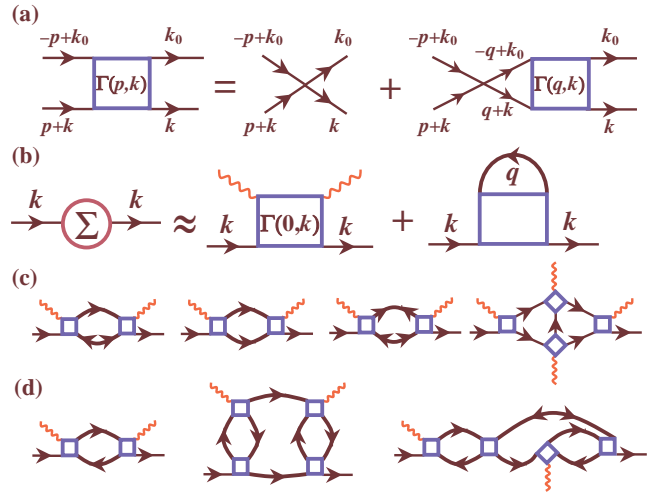}
\caption{(Color online.) (a) Equation for the vertex $\Gamma(p,k)$, where $k=(\omega,{\bf k})$. (b) Expression for the self-energy part $\Sigma(k)$ in the leading order in $\mu/H_c$ for $\bf k$ not very close to ${\bf k}_0$ and $(\pi,0)$. (c) and (d) Some higher order diagrams for $\Sigma(k)$ which are much smaller than those presented on slide (a) if $\mu\ll H_c$ and $\bf k$ is not very close to ${\bf k}_0$ and $(\pi,0)$. Lines with two arrows and wavy lines correspond to anomalous Green's functions and condensed particles, respectively. Diagrams the more simple of which are shown on slide (d) contribute to the leading correction to ${\rm Re}\Sigma(k)$ at ${\bf k}\sim(\pi,0)$.
\label{eq}}
\end{figure}

Formally Eq.~(\ref{hbg}) describes a dilute Bose gas with pare interaction between particles and with the chemical potential $\mu$. In 3D AF this equivalence leads to the solution at
$\mu\ll H_c$ \cite{bat84} by borrowing the results from the theory of the dilute Bose gas \cite{popov,belyaev}. Within this approach the long-range AF ordering in the plane perpendicular to the field appearing at $H<H_c$ corresponds to condensation of magnons with
momentum equal to ${\bf k}_0$. The density of particles is proportional to $\mu=H_c-H$. Therefore one can use results for dilute Bose gas when $\mu\ll H_c$. The ladder approximation for the vertex is valid and one has for the vertex the equation shown in Fig.~\ref{eq}(a). Normal $\Sigma(k)$ and anomalous $\Pi(k)$ self-energy parts are expressed via the vertex, where $k=(\omega,{\bf k})$ and the same notation is used below. In particular one has at $H<H_c$ for $\Sigma(k)$ the expression shown in Fig.~\ref{eq}(b) in the leading order in $\mu/H_c$ and corrections to it are small from other diagrams some of which are shown in Figs.~\ref{eq}(c) and \ref{eq}(d).

It is tempting to apply the same approach to 2D AF at $T=0$ but in contrast to 3D dilute Bose gas the theory of 2D one at $T=0$ can be developed only in the limit of exponentially small density of particles \cite{schick}. It means in the case of 2D AF that there is a good perturbation theory only for fields exponentially close to $H_c$, $\ln(H_c/\mu)\gg1$, while it is desirable to extend the theory to a more acceptable smallness of the parameter $\mu/H_c$. We show now that one can overcome this obstacle and construct a perturbation theory on $\mu/H_c$ to obtain the magnon spectrum for momenta $|{\bf k}-{\bf k}_0|\sim1$ in the leading order.

\begin{figure}
\centering
\includegraphics[scale=0.87]{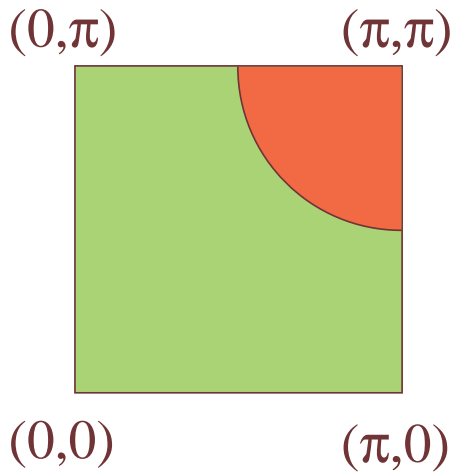}
\caption{(Color online.) A quarter of Brillouin zone is shown and an area around ${\bf k}_0$ is drawn in red in which the suggested perturbation approach does not work when $\ln(H_c/\mu)\sim 1$ (i.e., in which the condition \eqref{ineq} does not hold). The red area grows as the field reduces.
\label{bz}}
\end{figure}

We remind first that self-energy parts are small at $\mu\ll H_c$ and they vanish at $\mu=0$ \cite{popov,belyaev}. Then, one notices that for such $\bf k$ that the inequality
\begin{equation}
\label{ineq}
\epsilon_{\bf k}-\mu \gg |\Sigma(\epsilon_{\bf k},{\bf k})|
\end{equation}
satisfies, one can neglect $|\Pi(k)|^2$ in the GF denominator
$
{\cal D}(k) = (\omega - \epsilon_{\bf k} + \mu - \Sigma(k))(\omega + \epsilon_{\bf k} - \mu + \Sigma(-k)) + |\Pi(k)|^2
$
and the normal GF
$
G(k) = (\omega + \epsilon_{\bf k} - \mu + \Sigma(-k))/{\cal D}(k)
$
(see Refs.~\cite{popov,belyaev}) has the form
\begin{equation}
\label{gf}
G(k) = \frac{1}{\omega - (\epsilon_{\bf k}-\mu+\Sigma(k))}.
\end{equation}
It is implied in Eq.~\eqref{ineq} that $|\Sigma(\epsilon_{\bf k},{\bf k})|\sim|\Pi(\epsilon_{\bf k},{\bf k})|$ that is really the case in the entire BZ as particular estimations show. Inequality \eqref{ineq} is not satisfied only in an area around ${\bf k}_0$ that grows as the field reduces (see Fig.~\ref{bz}). Thus one has to calculate only normal self-energy part to obtain the magnon spectrum at $|{\bf k}-{\bf k}_0|\sim1$ (to be precise, when the condition \eqref{ineq} holds). 

It is the matter of direct estimations, similar to those carried out in Refs.~\cite{popov,belyaev} for dilute 3D Bose gas, to show using the smallness of $\Sigma$ and $\Pi$ that we have expression for $\Sigma(k)$ presented in Fig.~\ref{eq}(b) in the leading order in $\mu/H_c$ if $\bf k$ is not very close to ${\bf k}_0$ and $(\pi,0)$ (see Appendix for some details). Integrals in other diagrams, some of which are presented in Figs.~\ref{eq}(c) and \ref{eq}(d), become less singular at small $\mu$ at such external momenta as compared with cases of ${\bf k}\sim{\bf k}_0$ and $(\pi,0)$. As a result contribution from them estimated in the Appendix turns out to be of the next order in $\mu/H_c$. Diagrams, the more simple of which are shown in Fig.~\ref{eq}(d), should be also taken into account at ${\bf k}\sim(\pi,0)$ as we discuss below.

To calculate directly two diagrams drawn in Fig.~\ref{eq}(b) one needs to know the condensed particle density $\rho_0$ and the spectrum in the entire BZ which can be found only using
self-energy parts near ${\bf k}_0$ \cite{popov,belyaev}. The trick here bases on the fact that momenta of integration near ${\bf k}_0$ are essential in the second diagram in Fig.~\ref{eq}(b). Thus, one can put the momentum $q=(\omega_q,{\bf q})$ in the vertex equal to $k_0=(0,{\bf k}_0)$ and factor out the vertex. As a result we have
\begin{eqnarray}
\label{sen}
\Sigma(k) &=& 4n\Gamma(0,k),\\
\label{n} n &=& \rho_0 + \frac iN\sum_{{\bf q}\ne{\bf
k}_0}\int\frac{d\omega}{2\pi}e^{i\omega t}G(\omega,{\bf q}) =
\frac1N\sum_i\langle a_i^\dagger a_i\rangle,
\end{eqnarray}
where $t\to+0$ and $n$ is the density of particles which is expressed in spin-$\frac12$ AF via the magnetization $M=-\sum_i \langle S_i^z\rangle$ as $n = \frac12 - \frac MN$. The problem remains  to calculate $M$ but we can make use of results of previous numerical calculations.

Magnetization in spin-$\frac12$ square 2D AF in magnetic field was investigated before in the second order in $1/S$ \cite{zhitnik} and by numerical diagonalization of finite clusters \cite{magnum}. It was obtained in Ref.~\cite{zhitnik} that expression for the magnetization derived within the first order in $1/S$ fits the numerical data very good except for close vicinity of $H_c$. Thus we can approximate $n$ to high accuracy by its value obtained in the first order in $1/S$ for which one has after calculations following those of Ref.~\cite{zhitnik}
\begin{equation}
\label{nsw}
n = \frac{\tilde\mu}{2} + \frac{1-\tilde\mu}{N}
\sum_{\bf q} \frac{J_{\bf q}}{2J_{\bf 0}} \sqrt{\frac{J_{\bf 0} +
J_{\bf q}}{J_{\bf 0} + J_{\bf q} - 2J_{\bf
q}\tilde\mu(2-\tilde\mu)}},
\end{equation}
where $\tilde\mu=\mu/H_c$. In particular, one obtains from
Eq.~\eqref{nsw} $n\approx0.083$ and 0.048 for $H=0.9H_c$ and
$0.95H_c$, respectively. We conclude from Eq.~\eqref{nsw} that
$n\sim\mu/H_c$ if $H$ is not exponentially close to $H_c$.

It remains only to obtain the vertex $\Gamma(0,k)$. We imply for simplicity that there is exchange coupling between neighboring spins only with the exchange coupling constant $J$. It is convenient to try solution of the equation shown in Fig.~\ref{eq}(a) in the form 
\begin{equation}
\label{subst}
\Gamma(p,k) = 
\alpha(k) 
+ 
\beta(k)\sin\left(p_x + \frac12k_x\right) 
+ 
\gamma(k)\sin\left(p_z + \frac12k_z\right).
\end{equation}
Putting this expression in the equation one leads to a set of three linear algebraic equations on $\alpha(k)$, $\beta(k)$ and $\gamma(k)$ that can be readily solved. The result is quit cumbersome and we do not present it here. One leads to very compact expressions for the vertex at $p=0$ and $k=(\omega=\epsilon_{\bf k}+i\delta,{\bf k})$ in the particular cases of $|k_x|=|k_z|$ and $k_z=0$ having the form 
\begin{eqnarray}
\label{gdiag}
\Gamma(0,k) &=& 
\frac{J_{\bf 0}J_{({\bf k}+{\bf k}_0)/2}^2}{4{\cal T}J_{({\bf k}+{\bf k}_0)/2}^2-J_{\bf 0}^2} \quad\mbox{ for } \quad |k_x|=|k_z|,\\
\label{gkz}
\Gamma(0,k) &=& 
\frac{J_{\bf 0}(1-\cos k_x)}{4{\cal T}(1-\cos k_x) - 4}
\quad\mbox{ for } \quad k_z=0,
\end{eqnarray}
where 
\begin{equation}
\label{t}
{\cal T} = \frac1N \sum_{\bf q} 
\frac{J_{\bf 0}}{2(\epsilon_{\bf q} + \epsilon_{{\bf q}+{\bf k}+{\bf k}_0} - \epsilon_{\bf k} - i\delta)}. 
\end{equation}
We use these results below to calculate $\Sigma(k)$ for $\bf k$ directed along vectors ${\bf k}_0$ and $(\pi,0)$. As we discuss corrections to the spectrum in the leading order in $\mu/H_c$, we discard $\mu$ in the expressions for the vertex. 

\begin{figure}
\centering
\includegraphics[scale=0.87]{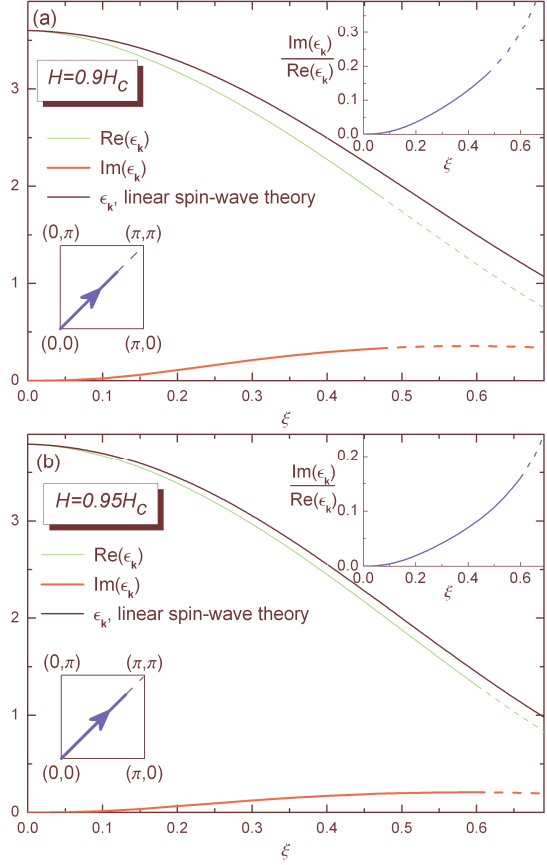}
\caption{(Color online.) Real and imaginary parts of magnon spectrum (in units of the exchange coupling constant $J$ between nearest spins) for ${\bf k}=\xi {\bf k}_0$ calculated in the leading order in $\mu/H_c$ for (a) $H=0.9H_c$ and (b) $H=0.95H_c$. Insets show ratios of imaginary and real parts of the spectrum. Dashed lines are for area in which the suggested perturbation technique does not work (inequality \eqref{ineq} does not hold).
\label{figspec}}
\end{figure}

It is difficult to calculate the vertex analytically in the entire BZ and we perform in the next section numerical integration to find it along some particular directions in BZ and for some particular $H$ values. 

\section{Results of the spectrum calculation}
\label{res}

Results for the momentum scan along ${\bf k}_0$ obtained using Eqs.~\eqref{gdiag} and \eqref{t} are plotted in Fig.~\ref{figspec} for $H=0.9H_c$ and $0.95H_c$. Solid (dashed) line in Fig.~\ref{figspec} is for the area in which inequality \eqref{ineq} does (does not) hold. Momenta ${\bf k}_c$ at which the solid line turns into the dashed one have been found from the assumption 
$
|\Sigma(\epsilon_{{\bf k}_c},{\bf k}_c)| = 0.2(\epsilon_{{\bf k}_c}-\mu)
$ 
that gives ${\bf k}_c\approx 0.47{\bf k}_0$ and $0.6{\bf k}_0$ for $H=0.9H_c$ and $0.95H_c$, respectively. Insets in Fig.~\ref{figspec} show that the ratio of the imaginary and the
real part of the spectrum does not exceed 0.17 upon increasing $|{\bf k}|$ up to $|{\bf k}_c|$. The spin-wave spectrum is also drawn in Fig.~\ref{figspec} obtained within the linear spin-wave theory (LSWT) and having the form 
$ 
\frac12\sqrt{(J_{\bf 0} + J_{\bf k})(J_{\bf 0} + (2(H/H_c)^2-1)J_{\bf k})} 
$. 
Notice that at ${\bf k}={\bf 0}$ the spectrum obtained within our approach is real and coincides with that derived in LSWT. This finding is in agreement with quite general arguments presented in Ref.~\cite{chub} according which quantum fluctuations do not change the classical spectrum at ${\bf k}={\bf 0}$ in quantum 2D AF at $H\ne0$. Quantum fluctuations shift down the spectrum at finite $\bf k$ as the first $1/S$ corrections do \cite{zhitcher99}.

\begin{figure}
\centering
\includegraphics[scale=0.93]{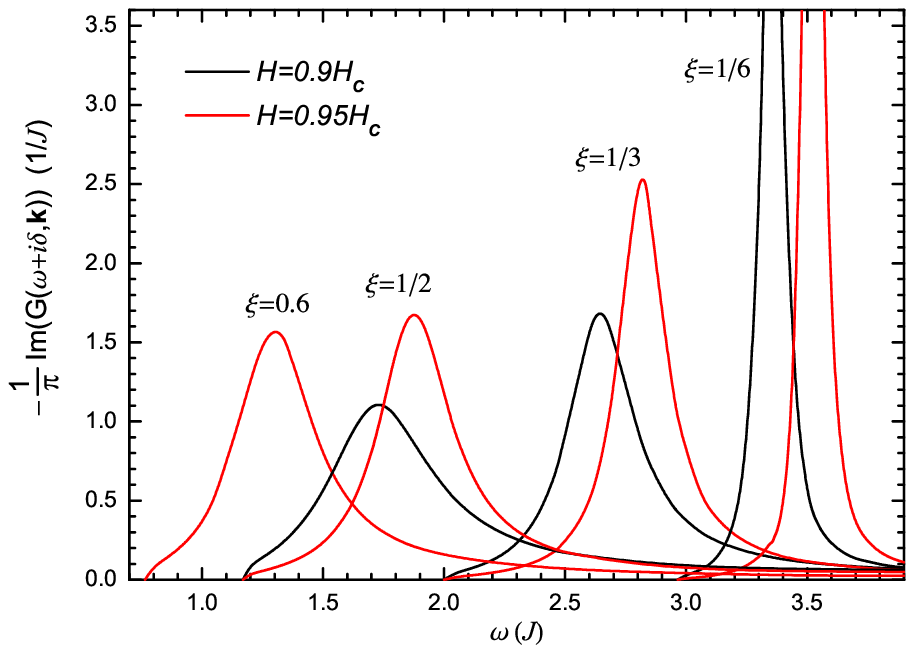}
\caption{(Color online.) Spectral function $-\frac1\pi{\rm Im}(G(\omega+i\delta,{\bf k}))$ at fixed momenta ${\bf k}=\xi {\bf k}_0$ (the value of $\xi$ is indicated near each couple of curves). Only the curve for $H=0.95H_c$ is shown for $\xi=0.6$.
\label{wscan}}
\end{figure}

Let us discuss the spectral function defined as $-\frac1\pi{\rm Im}G(\omega+i\delta,{\bf k})$. As is explained above, the vertex $\Gamma(0,k)$, where $k=(\omega,{\bf k})$, can be found exactly using representation \eqref{subst} with the following result for $|k_x|=|k_z|$ 
\begin{equation}
\label{sf}
\Gamma(0,k) = 
\frac{-(\omega -{J_0})  J_{0}^{2}  (2  {\cal T}  \omega +(1-2  {\cal T})  {J_0})+{J_0}  (-4 
{\cal T}  \omega +(-3+4  {\cal T})  {J_0})  J_{({\bf k}+{\bf k}_0)/2}^{2}-2  {\cal T}  J_{({\bf k}+{\bf k}_0)/2}^{4}}{2 
{J_0}  (2  {\cal T}  \omega   {J_0}+(1-2  {\cal T})  J_{0}^{2}-2  {\cal T}  J_{({\bf k}+{\bf k}_0)/2}^{2})},
\end{equation}
where
$
{\cal T} = 1/N \sum_{\bf q} 
J_{\bf 0}/2(\epsilon_{\bf q} + \epsilon_{{\bf q}+{\bf k}+{\bf k}_0}-\omega-i\delta).
$
These expressions can be simplified further at $\omega=\epsilon_{\bf k}$ and one leads to Eqs.~\eqref{gdiag} and \eqref{t}. We plot in Fig.~\ref{wscan} the spectral function for some momenta $\bf k\|{\bf k}_0$ obtained using Eq.~\eqref{sf}. One can see that curves look like peaks while peaks widths are quite large near the zone boundary. It should be noted that Fig.~\ref{wscan} is in contrast to the corresponding figure of Ref.~\cite{zhitcher99} for $H=0.9H_c$ obtained using $1/S$ expansion.

\begin{figure}
\centering
\includegraphics[scale=0.86]{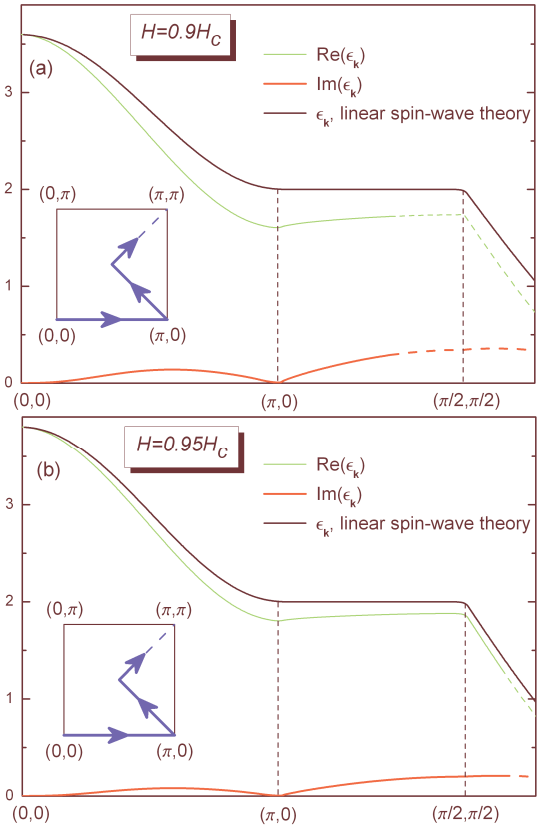}
\caption{(Color online.) Same as in Fig.~\ref{figspec} but along the path in BZ shown in insets. Rotonlike local minima are seen at $(\pi,0)$ accompanied by decrease of the damping near $(\pi,0)$.
\label{roton}}
\end{figure}

Results of the scan along another path in BZ are shown in Fig.~\ref{roton}. The most remarkable feature of this scan is the rotonlike minimum at $(\pi,0)$ similar to that obtained recently at $H=0$ \cite{roton1,roton2,christ,lumsden}. It should be noted that imaginary part of ${\cal T}$ diverges at ${\bf k}=(\pi,0)$ and the vertex vanishes because one has in Eq.~\eqref{t} 
$
\epsilon_{\bf q} + \epsilon_{{\bf q}+{\bf k}+{\bf k}_0}-\epsilon_{\bf k}
=
2J(1+\cos q_x)
$. 
Notice also that for ${\bf k}\|(\pi,0)$ the damping is at least of one order smaller than the real part of the spectrum even for $H=0.9H_c$. As is noticed above, extra diagrams should be taken into account at ${\bf k}\sim(\pi,0)$ the more simple of which are presented in Fig.~\ref{eq}(d) and which sum cannot be calculated analytically. It is shown in the Appendix that they give zero contribution to ${\rm Im}\Sigma(k)$ at ${\bf k} = (\pi,0)$ in the leading order and one concludes that there is a decrease of the damping near $(\pi,0)$. At the same time their real part contributes to the leading correction to ${\rm Re}\Sigma(k)$ at ${\bf k}\sim(\pi,0)$ being of the order of $\mu$. Then the real part of the spectrum can deviate near $(\pi,0)$ from curves drawn in Fig.~\ref{roton}. These diagrams are of different signs at ${\bf k}=(\pi,0)$ (e.g., the first two diagrams in Fig.~\ref{eq}(d) are negative and the last one is positive) and we cannot prove rigorously that there is the rotonlike minimum in the real part of the spectrum if $H$ is not very close to $H_c$. Meantime we demonstrate now using the theory of 2D dilute Bose gas \cite{schick} that the rotonlike minimum does exist if $\ln(H_c/\mu)\gg1$. One estimates in the leading order 
$
{\rm Re}\Sigma(k) \sim -\mu/\ln(H_c/\mu) + J\kappa^2\ln(H_c/\mu)
$ 
(the first and the second term here stem from the first diagrams shown in Fig.~\ref{eq}(d) and \ref{eq}(b), respectively) at ${\bf k} = (\pi,\kappa)$ and $\kappa\ll \mu/H_c$ while 
$
{\rm Re}\Sigma(k\sim\frac{k_0}{2})\sim\mu\ln(H_c/\mu)
$ 
and $\epsilon_{\bf k}\approx2J-J\kappa^2/2$. It seems to us likely that the minimum remains also at smaller $H$.

It is difficult to go beyond the first order in $\mu/H_c$ because, in particular, one needs to know self-energy parts near ${\bf k}_0$ to calculate the next order diagrams. We can estimate them to be ${\cal O}(\mu\sqrt{\mu/H_c})$ from the following consideration that works near $(\pi,0)$ only for ${\rm Im}\Sigma(k)$. Using relation $\Sigma(k_0)-\Pi(k_0)=\mu$ one obtains for the spectrum near AF vector $\sqrt{\epsilon_{\bf k}(\epsilon_{\bf k}+2\Pi(k_0))}$ \cite{popov,belyaev}. As a result diagrams, some of which are shown in Figs.~\ref{eq}(c) and \ref{eq}(d), are of the order of $\rho_0\sqrt{J\Pi(k_0)}$ or smaller (see Appendix for details). On the other hand one concludes from Eqs.~\eqref{n} and \eqref{nsw}, bearing in mind that the second term in Eq.~\eqref{n} is positive and of the order of $\Pi(k_0)$, that $\rho_0,\Pi(k_0)\sim\mu$ if $H$ is not exponentially close to $H_c$. 

Thus, one can expect from estimations by the order of magnitude that the approach discussed in the present paper works for $H>0.9H_c$. It should be noted that numerical evidences have appeared recently that short-wavelength magnons are unstable in some regions of BZ at $H\alt0.9H_c$. \cite{mc1,mc2} 

\section{Conclusion}
\label{con}

In conclusion, we discuss spin-$\frac12$ 2D Heisenberg AF on a square lattice at $T=0$ in strong magnetic field $H$ near its saturation value $H_c$. A perturbation approach is proposed to obtain spectrum of magnons with momenta not very close to AF vector ${\bf k}_0=(\pi,\pi)$ in the leading order in small parameter $\mu/H_c=(H_c-H)/H_c$. It is shown that only normal self-energy part $\Sigma(k)$ contributes to the spectrum renormalization for which we have two diagrams shown in Fig.~\ref{eq}(b). The sum of these diagrams is proportional to the number of particles which is expressed via uniform magnetization investigated before numerically (see Eqs.~\eqref{sen} and \eqref{n}). These diagrams are of the order of $\mu$. Higher-order diagrams some of which are shown in Fig.~\ref{eq}(c) and \ref{eq}(d) are of the order of ${\cal O}(\mu\sqrt{\mu/H_c})$ at $|{\bf k}-{\bf k}_0|\sim1$ except for the vicinity of the point ${\bf k}\sim (\pi,0)$, where diagrams the more simple of which are shown in Fig.~\ref{eq}(d) contribute to the leading part of ${\rm Re}\Sigma(k)$. We find that magnons are well-defined quasi-particles at $H>0.9H_c$ (see Figs.~\ref{figspec} and \ref{wscan}). A characteristic rotonlike local minimum in the spectrum is observed at ${\bf k}=(\pi,0)$ accompanied by decrease of the damping (see Fig.~\ref{roton}).

The approach suggested in the present paper can be used in discussion of short-wavelength excitations in 2D Bose gases of particles or quasi-particles. In the latter case one needs to calculate numerically the uniform magnetization to find the particle density $n$ in Eq.~\eqref{sen}. We note also that similar approach basing on the hard-core bosons formalism was proposed in Ref.~\cite{kot} for AFs with singlet ground state.

\begin{acknowledgments}

This work was supported by Russian Science Support Foundation, President of Russian Federation (grant MK-1056.2008.2), RFBR grant 07-02-01318, and Russian Programs "Quantum Macrophysics", "Strongly correlated electrons in semiconductors, metals, superconductors and magnetic materials" and "Neutron Research of Solids".

\end{acknowledgments}

\appendix

\section{Diagrams estimations}

This Appendix is devoted to estimation of diagrams. We demonstrate first that when $\bf k$ is not very close to ${\bf k}_0$ and $(\pi,0)$ the leading order corrections to the normal self-energy part $\Sigma(k)$ come from two diagrams shown in Fig.~\ref{eq}(b) and other diagrams are small some of which are shown in Figs.~\ref{eq}(c) and \ref{eq}(d). As it is explained in the main text, diagrams presented in Fig.~\ref{eq}(b) are of the order of $J\rho_0\sim\mu$. 

Let us discuss the first diagram in Fig.~\ref{eq}(d). Representing the normal GF as 
$
G(k) = (E(k)+\omega)/(\omega^2-\varepsilon_{\bf k}^2)
$,
where $E(k) = \epsilon_{\bf k}-\mu+\Sigma(-k)$ and $\varepsilon_{\bf k}$ is the renormalized spectrum, one estimates for the first diagram in Fig.~\ref{eq}(d) after integration over the frequency and putting $k=(\varepsilon_{\bf k},{\bf k})$
\begin{equation}
\label{loop}
	-8\rho_0\frac1N\sum_{\bf q}
	\frac{E(q)-\varepsilon_{\bf q}}{\varepsilon_{\bf q}}
	\frac{(E(q-k+k_0) + \varepsilon_{{\bf q} - {\bf k} + {\bf k}_0}) (\varepsilon_{\bf q} + \varepsilon_{{\bf q} - {\bf k} + {\bf k}_0})}{\varepsilon_{{\bf q} - {\bf k} + {\bf k}_0}[(\varepsilon_{\bf q} + \varepsilon_{{\bf q} - {\bf k} + {\bf k}_0})^2-\varepsilon_{\bf k}^2]}
	\Gamma(k-q,q)^2.
\end{equation}
One has in Eq.~\eqref{loop} when $q\sim k_0$ 
\begin{equation}
\label{fir}
	\frac{E(q)-\varepsilon_{\bf q}}{\varepsilon_{\bf q}}
	\approx 
	\frac{|\Pi(k_0)|^2}{(E(q)+\varepsilon_{\bf q})\varepsilon_{\bf q}} 
	\approx 
	\frac{|\Pi(k_0)|^2}{(\Pi(k_0)+\epsilon_{\bf k}+\varepsilon_{\bf q})\varepsilon_{\bf q}}, 
\end{equation}
where we use the relations \cite{popov,belyaev} $\Sigma(k_0)-\Pi(k_0)=\mu$ and 
$
\varepsilon_{\bf q} = \sqrt{\epsilon_{\bf q}(\epsilon_{\bf q}+2\Pi(k_0))}
$. 
Then, one has when $q\sim k_0$ and $\kappa\gg \Pi(k_0)/H_c$:
$
\varepsilon_{\bf q} 
\approx
\epsilon_{\bf q} + \Pi(k_0)
$
and 
\begin{equation}
\label{specdif}
	\varepsilon_{\bf q} + \varepsilon_{{\bf q} - {\bf k} + {\bf k}_0} - \varepsilon_{\bf k}
	\approx
	\frac J2 \left(\kappa^2 - (\kappa_x^2\cos k_x + \kappa_z^2\cos k_z) - 2(\kappa_x\sin k_x + \kappa_z\sin k_z) \right) + \Pi(k_0),
\end{equation}
where $\mbox{\boldmath $\kappa$}={\bf q} - {\bf k}_0$. One concludes from the above consideration bearing in mind that $\Sigma,\Pi\alt\mu$ that momenta of summation $\bf q$ near ${\bf k}_0$ are essential in Eq.~\eqref{loop}. 

Let us discuss the point ${\bf k} = (\frac\pi2,\frac\pi2)$. One has from Eq.~\eqref{specdif} in this case $\varepsilon_{\bf q} + \varepsilon_{{\bf q} - {\bf k} + {\bf k}_0} - \varepsilon_{\bf k}
\approx
-J(\kappa_x+\kappa_z)+\Pi(k_0)$.
As a result one obtains from Eqs.~\eqref{loop}, \eqref{fir}, and \eqref{specdif} that the first diagram shown in Fig.~\ref{eq}(d) is proportional to 
$
iJ\rho_0\sqrt{\mu/H_c}
\sim 
iJ(\mu/H_c)^{3/2}
$, 
where we use also that $\Pi(k_0)\sim\mu$, as it is explained in the main text. 

One leads to the same estimations of Eq.~\eqref{loop} for other $\bf k$ except for ${\bf k}\sim(\pi,0)$ and ${\bf k}\sim {\bf k}_0$. One has at ${\bf k}=(\pi,0)$ from Eq.~\eqref{specdif} $\varepsilon_{\bf q} + \varepsilon_{{\bf q} - {\bf k} + {\bf k}_0} - \varepsilon_{\bf k}
\approx J\kappa_x^2+\Pi(k_0)$. As a result the first diagram shown in Fig.~\ref{eq}(d) at ${\bf k}=(\pi,0)$ contributes to the main correction to ${\rm Re}\Sigma(k)$ being of the order of $J\rho_0\sim\mu$. Its contribution to ${\rm Im}\Sigma(k)$ is of the higher order in the parameter $\mu/H_c$. The same is true also for ${\bf k}\sim {\bf k}_0$ because we have at ${\bf k} = {\bf k}_0$: 
$
\varepsilon_{\bf q} + \varepsilon_{{\bf q} - {\bf k} + {\bf k}_0} - \varepsilon_{\bf k} 
=
2\varepsilon_{\bf q} 
\approx 
J\kappa^2+\Pi(k_0)
$.

We assume in the above estimations that $\Gamma(k-k_0,k_0)\sim J$ in Eq.~\eqref{loop}. It is really the case as the results show of particular calculations of $\Gamma(k-k_0,k_0)$ carried out as it is described in the main text. Meantime the vertex $\Gamma(0,k)\ll J$ in the vicinity of the point ${\bf k}=(\pi,0)$ because $\cal T$ diverges (see the main text). Thus only diagrams containing $\Gamma(k-k_0,k_0)$ contribute to the leading correction to ${\rm Re}\Sigma(k)$ at ${\bf k}\sim(\pi,0)$. The more simple of such diagrams are shown in Fig.~\ref{eq}(d).

\bibliography{damref}

\end{document}